\documentclass[graybox]{svmult}

\usepackage[section]{placeins} 
\usepackage{amssymb}
\usepackage{amsmath}
\usepackage{upgreek}
\usepackage{pdflscape}
\usepackage{listings}
\usepackage{multirow}
\usepackage[percent]{overpic}
\usepackage{multicol}
\usepackage{color}
\usepackage{stmaryrd}
\usepackage{xfrac}
\usepackage[capitalise]{cleveref}
\usepackage{booktabs}
\usepackage[section]{placeins} 
\usepackage[section]{algorithm}
\usepackage{calc}
\usepackage[titletoc]{appendix}
\usepackage{siunitx}
\usepackage{mathtools}
\usepackage[sort&compress,square,numbers]{natbib}

\usepackage{tikz-dimline}
\usepackage{graphicx}
\usepackage{graphics}
\usepackage{wrapfig}
\usepackage{float}
\usepackage{subfig}
\usepackage[percent]{overpic}
\usepackage{varwidth}
\usepackage{tikz}
\usepackage{pgfplots}
\usepackage{adjustbox}
\usetikzlibrary{arrows,matrix,positioning,fit}
\usetikzlibrary{shapes,positioning}
\usetikzlibrary{backgrounds}
\usepackage{tikz-layers}
\usepgfplotslibrary{fillbetween}
\usetikzlibrary{intersections}
\pgfplotsset{compat=1.14}
\usepgfplotslibrary{colorbrewer}
\usepgfplotslibrary{patchplots}
\usepgfplotslibrary[colorbrewer]
\usetikzlibrary{pgfplots.colorbrewer}
\usetikzlibrary[pgfplots.colorbrewer]
\usepgfplotslibrary{units}
\usetikzlibrary{spy}
\usepackage{pgfplotstable}
\usepackage{arrayjobx}
\graphicspath{ {./figs/} }
\usetikzlibrary{external}

\newlength\myheight
\newlength\mydepth
\settototalheight\myheight{Xygp}
\newcommand*\inlinegraphics[1]{%
  \settototalheight\myheight{Xygp}%
  \settodepth\mydepth{Xygp}%
  \raisebox{-\mydepth}{\includegraphics[height=\myheight]{#1}}%
}
\newcommand\orcid[1]{\href{https://orcid.org/#1}{\inlinegraphics{orcid_16x16.png}}}

\makeatletter
\def\BState{\State\hskip-\ALG@thistlm}
\makeatother

\usepackage{type1cm}        
\usepackage{makeidx}         
\usepackage{graphicx}        
\usepackage{multicol}        
\usepackage[bottom]{footmisc}

\usepackage{newtxtext}       %
\usepackage[varvw]{newtxmath}       
\usepackage{eso-pic}
\newcommand\AtPageUpperMyright[1]{\AtPageUpperLeft{%
 \put(\LenToUnit{0.33\paperwidth},\LenToUnit{-1cm}){%
     \parbox{0.8\textwidth}{\raggedleft\fontsize{9}{11}\selectfont #1}}%
 }}%
\newcommand{\conf}[1]{%
\AddToShipoutPictureBG*{%
\AtPageUpperMyright{#1}
}
}

\makeindex             

\begin{document}

\title*{Towards full molecular gas dynamics simulations of complex flows via the Boltzmann equation}
\titlerunning{Full molecular gas dynamics simulations of complex flows via the Boltzmann eqn.}
\author{Tarik Dzanic and Luigi Martinelli}

\institute{Tarik Dzanic \at Princeton University, Princeton, NJ, 08544, USA, \email{tdzanic@princeton.edu}
\and Luigi Martinelli \at Princeton University, Princeton, NJ, 08544, USA, \email{martinel@princeton.edu}}
%
%
\maketitle
\conf{CUFS-2024-10, Cambridge, United Kingdom, 4-5 March 2024} 

\abstract{This work explores the capability of simulating complex fluid flows by directly solving the Boltzmann equation. Due to the high-dimensionality of the governing equation, the substantial computational cost of solving the Boltzmann equation has generally limited its application to simpler, two-dimensional flow problems. Utilizing a combination of high-order spatial discretizations and discretely-conservative velocity models along with their highly-efficient implementation on massively-parallel GPU computing architectures, we demonstrate the current ability of directly solving the Boltzmann equation augmented with the BGK collision model for complex, three-dimensional flows. Numerical results are presented for a variety of these problems including rarefied microchannels, transitional and turbulent flows, and high-speed atmospheric re-entry vehicles, showcasing the ability of the approach in accurately predicting complex nonlinear flow phenomena and non-equilibrium effects.}

\section{Introduction}
For the vast majority of problems in fluid dynamics, numerical techniques have broadly relied on the solution of the Navier--Stokes equations to give insight into the dynamics of fluid flows. Embedded in these governing equations is the assumption that the fluid can be treated as continuum, allowing for the use of the macroscopic conservation laws (i.e., conservation of density, momentum, and energy) to characterize the flow. However, in certain applications ranging from microflows to hypersonic aeronautics, this assumption can start to break down as the flow begins to experience strong thermodynamic non-equilibrium effects, such that it is necessary to revert to more general governing equations derived from the kinetic theory of gases. These kinetic schemes can offer a more robust approach which can accurately describe these complex flow physics while seamlessly recovering the hydrodynamic equations in the continuum limit.

In this work, we explore the capability of simulating complex fluid flows by directly solving the Boltzmann equation for molecular gas dynamics which underpins the macroscopic behavior of the flow. In this approach, the flow is simply represented through the evolution of a scalar particle distribution function from which a macroscopic flow state can be recovered that is equally valid across both the rarefied and continuum flow regimes. While this approach can be highly advantageous, owing both to its generalizability and the numerical simplicity of the primarily linear governing equation, its application to complex fluid flows has been very limited primarily as a result of the substantial computational cost of directly solving the Boltzmann equation. This cost can be attributed to two main sources, dimensionality and collision modeling. For the former, the phase space in the governing equation, which represents a probability density of a particle existing at some location with some velocity, can require discretizations of up to six dimensions which results in a rapidly increasing computational cost with respect to resolution. For the latter, the approximation of the particle collision process can be extremely computationally intensive, requiring the numerical evaluation of integrals of even higher dimensionality. As a result, directly solving the Boltzmann equation has typically been limited to simpler, often two-dimensional, flow problems, and the extension to more complex three-dimensional flows has generally been considered to be computationally intractable.  

While the computational cost of solving the Boltzmann equation can be very restrictive, the approach may present some unique benefits which motivate its development and application. Aside from its typical use in rarefied gas dynamics, for which only kinetic descriptions of the flow are truly valid, the use of the Boltzmann equation in other flow regimes can potentially lead to enhancing our understanding of fundamental flow flows and offer more robust and accurate approaches for complex flow problems. For flows in the low-speed continuum regime, where the hydrodynamic equations can accurately describe the flow, the solution of the Boltzmann equation encodes the flow dynamics through the linear evolution of a high-dimensional particle distribution function which represents the flow in a manner that is inaccessible from the solution of the hydrodynamic equations. As such, the Boltzmann equation for these flows can offer a radically different perspective for analyzing fundamental flow problems such as transition to turbulence and present opportunities for alternate approaches to turbulence modeling. Furthermore, the Boltzmann equation can offer a robust approach for simulating the multi-scale nature of high-speed flows in re-entry and hypersonic conditions as it can accurately approximate the flow physics across the entire range of flow regimes encountered and may provide a better framework for simulating the complex high-temperature aerothermodynamic effects which are encountered in these applications.

To this end, this work presents an overview of some developments introduced by the authors for drastically reducing the computational cost of solving the Boltzmann equation, enabling its application to complex three-dimensional flow problems that were previously intractable. These improvements can broadly be categorized by three separate advancements: 1) the use of high-order spatial discretizations which allow for higher-fidelity approximations that reduce the required resolution; 2) their combination with discretely-conservative collision models which avoid the large computational cost of directly computing the collision operator and memory requirements associated with ensuring macroscopic conservation; and 3) the efficient implementation of these numerical methods on modern massively-parallel GPU computing architectures. A brief summary of this numerical approach is presented in \cref{sec:methodology}, after which an overview of numerical results obtained by this approach is shown \cref{sec:results}. Conclusions are then drawn in \cref{sec:conclusion} along with a discussion on potential future development and applications. 
\section{Methodology}\label{sec:methodology}
The Boltzmann equation can be represented as a scalar conservation law with a linear transport term and nonlinear source term, given as
\begin{equation}\label{eq:boltzmann}
    \partial_t f (\mathbf{x}, \mathbf{u}, t) + \mathbf{u} {\cdot} \nabla f = \mathcal C(f, f'),
\end{equation}
where $\mathbf{x} \in \Omega^{\mathbf{x}}$ is the physical space in a physical domain $\Omega^{\mathbf{x}} \subseteq \mathbb R^d$ for some spatial dimension $d$, $\mathbf{u} \in \Omega^{\mathbf{u}}$ is the associated velocity space in a velocity domain $\Omega^{\mathbf{u}} \in \mathbb R^m$ for some velocity dimension $m \geq d$, $f (\mathbf{x}, \mathbf{u}, t)\in \mathbb R$ is a scalar particle distribution function, and $\mathcal C(f, f')$ is the collision operator that models the effects of particle interactions \citep{Cercignani1988}. The distribution function represents a phase space probability density for a particle existing at a given location $\mathbf{x}$ with a given velocity $\mathbf{u}$. From this distribution function, the conserved flow variables $\mathbf{Q}(\mathbf{x}, t)$ can be recovered through its moments as
\begin{equation}\label{eq:moments}
    \mathbf{Q}(\mathbf{x}, t) = \left[\rho, \rho \mathbf{U}, E \right]^T = 
    \int_{\mathbb R^m} f (\mathbf{x}, \mathbf{u}, t)\ \boldsymbol{\psi} (\mathbf{u}) \ \mathrm{d}\mathbf{u},
\end{equation}
where $\rho$ is the density, $\rho \mathbf{U}$ is the momentum vector, $E$ is the total energy, and $\boldsymbol{\psi} (\mathbf{u}) \coloneqq [1, \mathbf{u}, (\mathbf{u}\cdot\mathbf{u})/2]^T$ is the vector of collision invariants. 

To reduce the cost of directly computing the collision operator, an approximate collision model introduced by \citet{Bhatnagar1954}, known as the Bhatnagar--Gross--Krook (BGK) model, was instead used. For this model, collision is approximated as a relaxation process to thermodynamic equilibrium, represented as
\begin{equation}
    C(f, f') \approx \frac{g(\mathbf{x}, \mathbf{u}, t) - f(\mathbf{x}, \mathbf{u}, t)}{\tau},
\end{equation}
where $g(\mathbf{x}, \mathbf{u}, t)$ is the local thermodynamic equilibrium distribution and $\tau$ is the relaxation time scale. For a monatomic particle, the equilibrium distribution function is given by the Maxwellian 
\begin{equation}
    g(\mathbf{x}, \mathbf{u}, t) = \frac{\rho (\mathbf{x}, t)}{\left[2 \pi \theta(\mathbf{x}, t) \right]^{d/2}}\exp \left [-\frac{ \|\mathbf{u} - \mathbf{U}(\mathbf x, t) \|_2^2}{2 \theta (\mathbf x, t)} \right],
\end{equation}
where $P = (\gamma - 1)(E - \frac{1}{2}\rho \mathbf{U}\cdot\mathbf{U})$ is the pressure and $\theta = P/\rho$ is a scaled temperature, and the collision time is given by
\begin{equation}
    \tau = \frac{\mu}{P},
\end{equation}
where $\mu$ is the dynamic viscosity which can be set adaptively based on some temperature-based viscosity law (e.g., power law, Sutherland's law). 

\subsection{Discretization}
\begin{figure}[tbhp]
    \begin{centering}
    \adjustbox{width=0.6\linewidth, valign=b}{\input{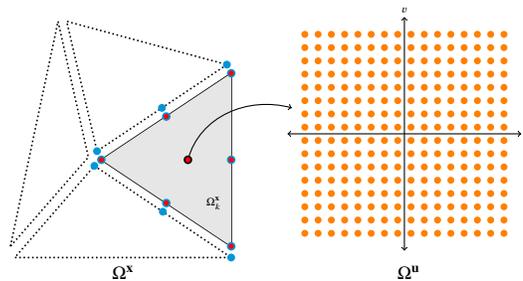}}
    \caption{\label{fig:scheme} Schematic of a two-dimensional phase space discretization using an unstructured spatial domain $\Omega^{\mathbf{x}}$ with $\mathbb P_2$ elements and a velocity domain $\Omega^{\mathbf{u}}$ with $N_v = 16^2$. Circles denote the spatial solution nodes (red), interface flux nodes (blue), and velocity space nodes (orange), respectively.}
    \end{centering}
\end{figure}

We present here a very brief overview of the numerical approach used to solve the Boltzmann--BGK equation. For a more in-depth description, the reader is referred to \citet{Dzanic2023b} and \citet{Dzanic2023}. We utilize a nodal spatial and velocity discretization, where each degree of freedom corresponds to a given location in the spatial and velocity domains, i.e.
\begin{equation}
    f_{ij} = f(\mathbf{x}_i, \mathbf{u}_j).
\end{equation}
A schematic of this discretization is shown in \cref{fig:scheme}. In this approach, the transport (spatial) and collision (velocity) terms are decoupled, resulting in a separate linear advection equation for each velocity node $\mathbf{u}_j$ with a spatially-independent nonlinear source term for each spatial node $\mathbf{x}_i$.

\textit{Spatial discretization.} To approximate the particle transport portion of the Boltzmann equation, we utilize high-order discontinuous spectral element methods \citep{Hesthaven2008DG}, specifically the flux reconstruction scheme of \citet{Huynh2007}, to discretize the spatial domain instead of standard low-order finite volume methods. These numerical schemes, which possess the geometric flexibility of finite volume methods while retaining the arbitrarily high-order accuracy and efficiency of finite difference methods, can achieve the equivalent accuracy of highly-resolved low-order schemes with substantially fewer degrees of freedom, drastically reducing the computational cost and memory requirements of accurately resolving complex fluid flows. 

In this approach, the solution within each element $\Omega_k^{\mathbf{x}}$ is represented by a high-order interpolating polynomial across a set of $N_s$ solution nodes $\mathbf{x}_i^s$ as
\begin{equation}
    f (\mathbf{x}) = \sum_{i = 1}^{N_s} f (\mathbf{x}^s_i) {\phi}_i (\mathbf{x}),
\end{equation}
where ${\phi}_i (\mathbf{x})$ is the nodal interpolating polynomial associated with the given solution node. Without loss of generality, we present this spatial discretization in terms of an arbitrary velocity node $\mathbf{u}_0$.

The flux is calculated via flux reconstruction methodology as a collocation projection with an interface correction term as 
\begin{equation}
    \mathbf{F}(\mathbf{x}) = \mathbf{u}_0 f (\mathbf{x}) + \sum_{i = 1}^{N_f} \left[F^I_i - \mathbf{u}_0\cdot \mathbf{n}_i f (\mathbf{x}^f_i) \right] \mathbf{g}_i (\mathbf{x}),
\end{equation}
where $\mathbf{x}^f_i$ is a set of $N_f$ interface flux nodes, $\mathbf{n}_i$ is their associated outward-facing normal vector, and $\mathbf{g}_i$ is their associated correction function which is chosen to recover the nodal discontinuous Galerkin approach \citep{Huynh2007, Hesthaven2008DG, Trojak2021}. Furthermore, $F^I_i$ is the upwind-biased common interface flux, i.e., 
\begin{equation}
    F_i^I = \begin{cases}
    u_n f_i^-, \quad \quad \mathrm{if} \ u_n > 0,\\
    u_n f_i^+, \quad \quad \mathrm{else},
    \end{cases}
\end{equation}
where $u_n = \mathbf{u}_0 \cdot \mathbf{n}_i$ and the superscripts $-$ and $+$ denote the interior value and the exterior value of the solution at the interface, respectively. 

As these high-order schemes do not typically preserve a maximum principle, they do not guarantee that the distribution function remains strictly positive which is physically inconsistent for a probability measure and can result in the divergence of the numerical scheme as it may not ensure a strictly-positive macroscopic density and temperature. Therefore, the spatial discretization is augmented with the high-order, positivity-preserving limiter of \citet{Zhang2010}, which contracts the solution to the element-wise mean $\bar{f}$ if the distribution function attains a negative value at any spatial node, i.e.,
\begin{equation}
    \hat{f}(\mathbf{x}_i) = \bar{f} + \beta\left[ f(\mathbf{x}_i) - \bar{f} \right],
\end{equation}
where
\begin{equation}
    \beta = \min \left [ \left | \frac{\bar{f}}{\bar{f} - f^{\min}}\right |, 1\right] \quad \mathrm{and} \quad  f^{\min} = \min f(\mathbf{x}_i)\ \forall \ i \in \{1, ..., N_s\}.
\end{equation}
As a result of this limiting procedure, the positivity of the distribution function and, by extension, the macroscopic density and temperature, are ensured while retaining the high-order accuracy of the underlying numerical scheme.

\textit{Velocity discretization.} To approximate the particle collision portion of the Boltzmann equation, the
velocity space is represented as a finite subset of the infinite velocity domain $\Omega^{\mathbf{u}} \subset \mathbb R^m$, which is discretized by a uniform Cartesian grid of $N_v$ velocity nodes, shown on the right-hand side of \cref{fig:scheme}. The extent of the velocity domain is taken as some factor of the maximum thermal velocity in the problem, where a factor of 4 results in $99\%$ of the distribution function being contained within $\Omega^{\mathbf{u}}$ for near-Maxwellian distributions. 

To evaluate the equilibrium distribution function $g$ corresponding to $f$, it is necessary to compute the moments of the $f$. This is approximated via a discrete integration operator $\mathbf{M}$ with strictly-positive entries, i.e.,
\begin{equation}
     \mathbf{M} \cdot \mathbf{f} \approx \int_{\mathbb R^m} f (\mathbf{u}) \ \mathrm{d}\mathbf{u}, \quad \quad \mathbf{Q} =  \mathbf{M} \cdot \left[\mathbf{f} \otimes \boldsymbol{\psi} \right].
\end{equation}
As the nodes in the velocity domain are uniformly distributed and the distribution functions are smooth and compactly supported, the spectral convergence of the trapezoidal rule makes it an ideal choice for quadrature \citep{Trefethen2014}. However, since this integration operator cannot exactly represent the integration of the moments, these integration errors result in discrete conservation errors in the solution as the distribution function the solution is relaxing to via the BGK operator does not possess the same macroscopic state, i.e.,
\begin{equation}
    \mathbf{M} \cdot \left[\mathbf{f} \otimes \boldsymbol{\psi} \right]  \neq \mathbf{M} \cdot \left[\mathbf{g} \otimes \boldsymbol{\psi} \right].
\end{equation}
While these conservation errors can be contained to a reasonable tolerance with increased resolution in the velocity domain, it was shown in \citet{Dzanic2023} and \citet{Dzanic2023b} that the primary source of approximation error in the scheme stemmed from the conservation error and that if discrete conservation could be ensured, accurate approximation of particle collision could be obtained with much fewer degrees of freedom in the velocity domain. 

This discretely-conservative velocity model was achieved through the discrete velocity model (DVM) approach of \citet{Mieussens2000}. In this approach, a discrete equilibrium distribution function is sought which satisfies the discrete compatibility condition (i.e., possesses the same moments as the distribution function) and satisfies the discrete form of Boltzmann's H-theorem. It was shown in \citet{Mieussens2000} that this discrete distribution function is represented by a Maxwellian formed around a perturbed macroscopic state $\mathbf{Q}'$ which converges to the true macroscopic state $\mathbf{Q}$ in the limit of infinite velocity resolution. However, as there does not exist a closed-form expression for this perturbed state, it must be computed numerically via a nonlinear optimization process. Due to the low dimensionality of the optimization problem and the presence of a closed-form expression of the Jacobian, a discretely-conservative approximation of the BGK operator could be achieved easily and efficiently with as few as two iterations of Newton's method. For an in-depth overview of this numerical approach, the reader is referred to \citet{Dzanic2023}, Section 3.4. As a result of this discrete conservation property, it was shown that the required resolution for accurately approximating complex fluid flows could be decreased by up to two orders of magnitude \citep{Dzanic2023, Dzanic2023b}, requiring as few as eight velocity nodes per dimension.

\subsection{Implementation}

The presented numerical approach was implemented within PyFR \citep{Witherden2014}, a high-order flux reconstruction solver which can efficiently target massively-parallel GPU computing architectures. Due to the space-velocity decoupling of the transport and collision operators and the compute intensive nature of the discrete velocity model, the numerical scheme could be very efficiently parallelized for GPU computing. As the purpose of this numerical approach is to target spatially well-resolved flows tending towards direct numerical simulation where the collision operator is not excessively stiff compared to the transport term, the use of explicit time stepping schemes was deemed preferable due to the many computational advantages they offer for GPU computing. 

For the numerical results to be presented, computations were performed on up to 80 32 GiB NVIDIA V100 GPUs with up to approximately $10^{11}$ total degrees of freedom. The numerical experiments indicate that for three-dimensional flows, the total computational cost for simulating flows via the Boltzmann equation is roughly one to two orders of magnitude higher than an equivalently-resolved Navier--Stokes simulation \citep{Dzanic2023b}. While this additional cost is somewhat substantial, these algorithmic improvements still make it entirely feasible to simulate complex fluid flows on modern compute hardware. In comparison, standard low-order finite volume approaches with nodal velocity models would require roughly an additional two orders of magnitude more computational effort for an equivalent level of resolution. 
\section{Results}\label{sec:results}
\begin{figure}[h!]
    \centering
    \subfloat[$Kn = 0.01$] {
    \adjustbox{width=0.33\linewidth,valign=b}{\includegraphics[width=\textwidth]{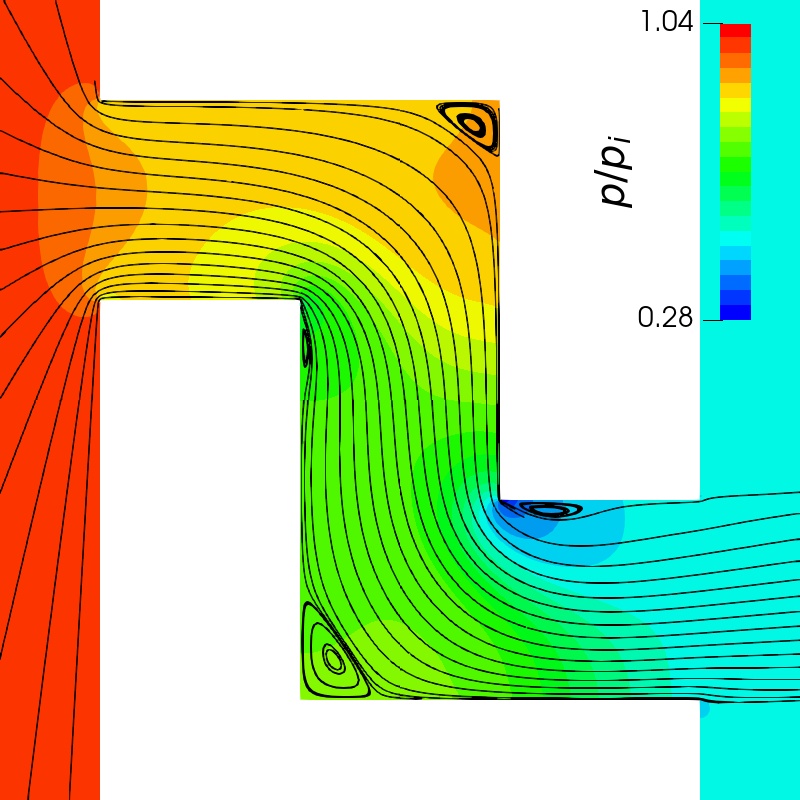}}}
    \subfloat[$Kn = 0.02$] {
    \adjustbox{width=0.33\linewidth,valign=b}{\includegraphics[width=\textwidth]{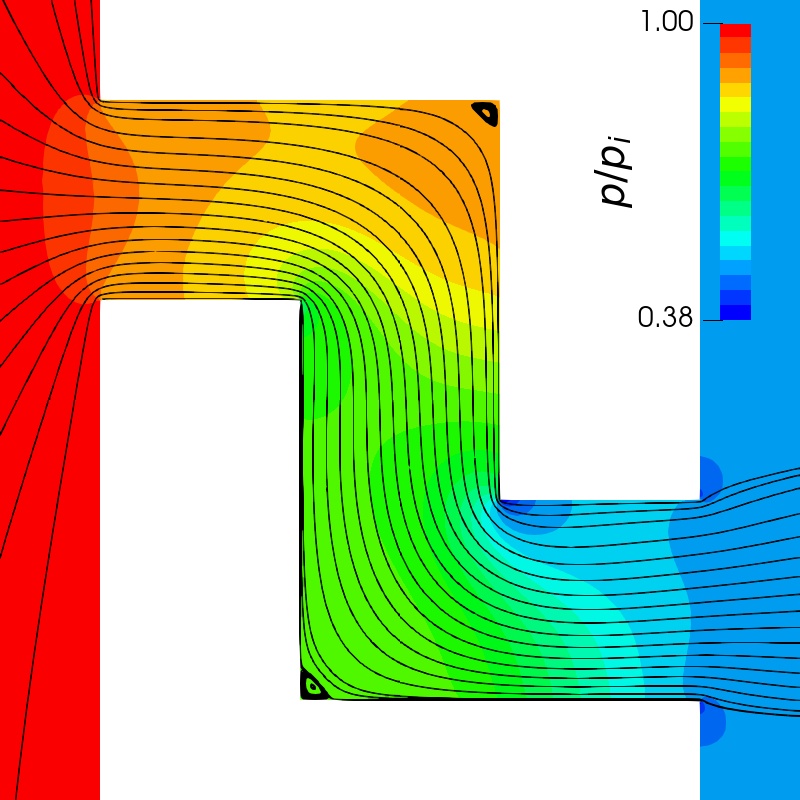}}}
    \subfloat[$Kn = 0.05$] {
    \adjustbox{width=0.33\linewidth,valign=b}{\includegraphics[width=\textwidth]{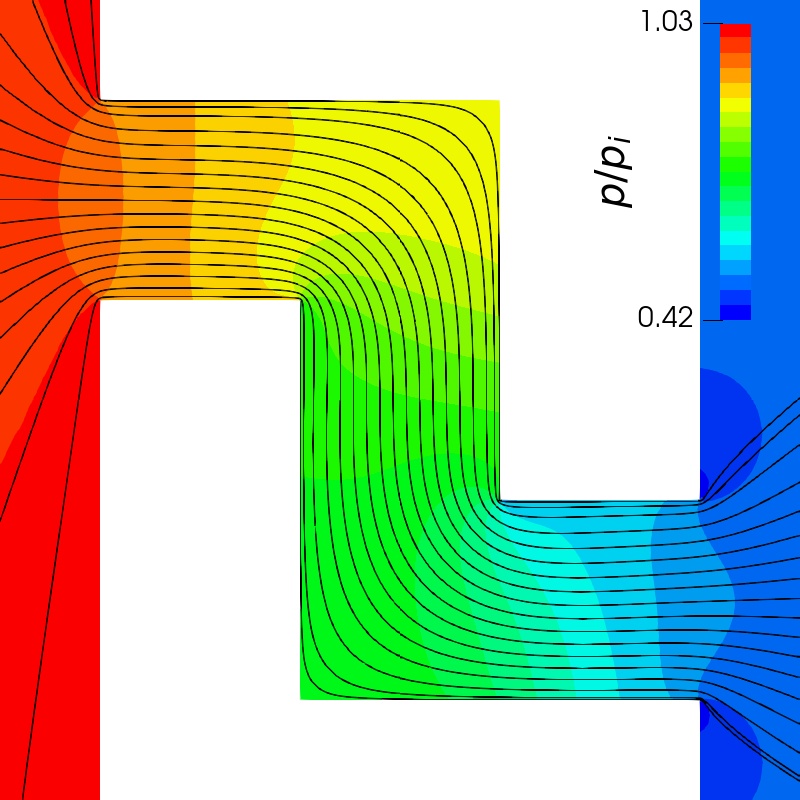}}}
    \caption{\label{fig:microchannel} Contours of normalized pressure overlaid with velocity streamlines for the bent microchannel problem at $Kn = 0.01$ (left), $Kn = 0.02$ (middle), and $Kn = 0.05$ (right) computed using a $\mathbb P_3$ approximation.}
\end{figure}

In this section, an overview of selected numerical experiments previously presented by the authors \citep{Dzanic2023, Dzanic2023b} is shown along with some novel results. A common application of the Boltzmann equation (and kinetic schemes in general) is for microchannels, where the small characteristic length scales result in very low Reynolds number flows, yielding strong rarefaction and non-equilibrium effects which cannot be accurately modeled using the standard Navier--Stokes equations. These geometries and flow conditions are commonly encountered in biomedical engineering applications and micro-electro mechanical system (MEMS) devices. One of the validation studies of the numerical approach was performed for the rarefied bent microchannel case of \citet{Ho2020}. A subset of the results of the numerical experiment is shown in \cref{fig:microchannel} as contours of pressure with velocity streamlines at varying Knudsen numbers. Non-equilibrium effects can be observed as the increasing Knudsen number results in higher degrees of slip velocity at the wall, affecting the structure of the flow around the concave and convex corners. These results showed excellent agreement with the reference data and showcased how a discretely-conservative velocity model can accurately resolve non-equilibrium effects with very few degrees of freedom. For example, in this experiment, a resolution of $N_v = 8^2$ was found to be sufficient to be converged in the velocity domain. At this resolution, the extension to three-dimensional microchannels was performed with only a factor of three increase in the computational cost in comparison to a standard Navier--Stokes approach.

\begin{figure}[h!]
    \centering
    \subfloat[Enstrophy]{\adjustbox{width=0.48\linewidth, valign=b}{\begin{tikzpicture}[spy using outlines={rectangle, height=3cm,width=2.5cm, magnification=3, connect spies}]
    \begin{axis}
    [
        axis line style={latex-latex},
        axis y line=left,
        axis x line=left,
        clip mode=individual,
        xmode=linear, 
        ymode=linear,
        xlabel = {$t$},
        ylabel = {$\varepsilon$},
        xmin = 0, xmax = 20,
        ymin = 0.00, ymax = 0.015,
        legend cell align={left},
        legend style={font=\scriptsize, at={(1.0, 1.0)}, anchor=north east},
        ytick = {0,0.004,0.008,0.012,0.016},
        x tick label style={/pgf/number format/.cd, fixed, fixed zerofill, precision=0, /tikz/.cd},
        y tick label style={/pgf/number format/.cd, fixed, fixed zerofill, precision=1, /tikz/.cd},	
        scale = 0.85
    ]
        
        \addplot[color=gray, style={thick}, only marks, mark=o, mark options={scale=0.8}, mark repeat = 3, mark phase = 0] table[x=t, y=enst, col sep=comma, mark=*]{./figs/data/tgvNS_256_post.csv};
        \addlegendentry{Reference};
        
        \addplot[color=black, style={very thick, dotted}] table[x=t, y=enst, col sep=comma]{./figs/data/tgvNS_192_post.csv};
        \addlegendentry{Navier--Stokes};
        
        \addplot[color=red!80!black, style={thick}] table[x=t, y=enst, col sep=comma]{./figs/data/tgvBGK_192_post.csv};
        \addlegendentry{Boltzmann--BGK};
    \end{axis}

\end{tikzpicture}}}
    ~
    \subfloat[Kinetic energy spectra]{\adjustbox{width=0.48\linewidth, valign=b}{\begin{tikzpicture}[spy using outlines={rectangle, height=3cm,width=2.3cm, magnification=3, connect spies}]
	\begin{loglogaxis}[name=plot1,
		xlabel={$k$},
		xmin=1,xmax=256,
		ylabel={$E(k)$},
		ymin=1e-4,ymax=1e-2,
		legend style={at={(0.03,0.03)},anchor=south west,font=\small},
		legend cell align={left},
		style={font=\normalsize},,	
        scale = 0.85
        ]

        \addplot[ color=gray, style={thick}] table[x=k, y=E, col sep=comma, mark=* ]{./figs/data/tgvNS_256_spectra.csv};
        \addlegendentry{Reference};
        
        \addplot[color=black, style={very thick, dotted}] table[x=k, y=E, col sep=comma]{./figs/data/tgvNS_192_spectra.csv};
        
        \addplot[color=red!80!black, style={thick}] table[x=k, y=E, col sep=comma]{./figs/data/tgvBGK_192_spectra.csv};
			
		\addplot[color=gray, style={dashed, thick},forget plot] coordinates{(50, 1e-2) (256, 0.00065748128)};
	    \node [below,color=black] at (axis cs:100,.008) {$-\frac{5}{3}$};
		\end{loglogaxis} 		
	\end{tikzpicture}}}
    \caption{\label{fig:tgv_profiles} Dissipation measured by enstrophy (left) and turbulent kinetic energy spectra at $t = 10$ (right) for the compressible Taylor-Green vortex at $Re = 1600$ computed using a $\mathbb P_3$ approximation.}
\end{figure}
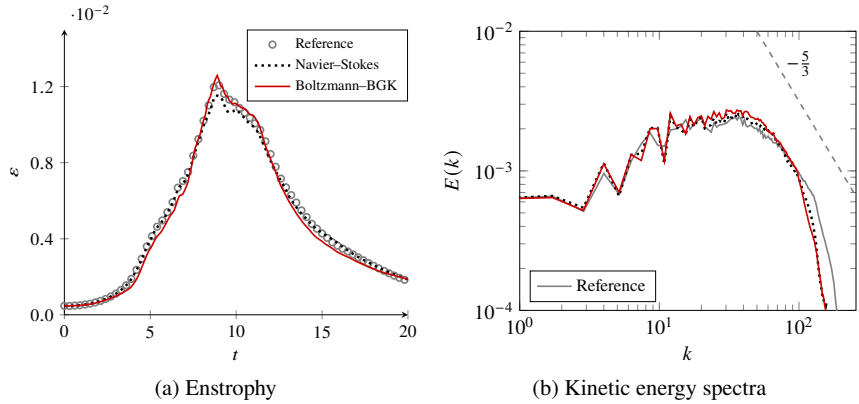

\begin{figure}[h!]
    \centering
    \subfloat[Navier--Stokes]{\adjustbox{width=0.48\linewidth, valign=b}{\includegraphics[width=\textwidth]{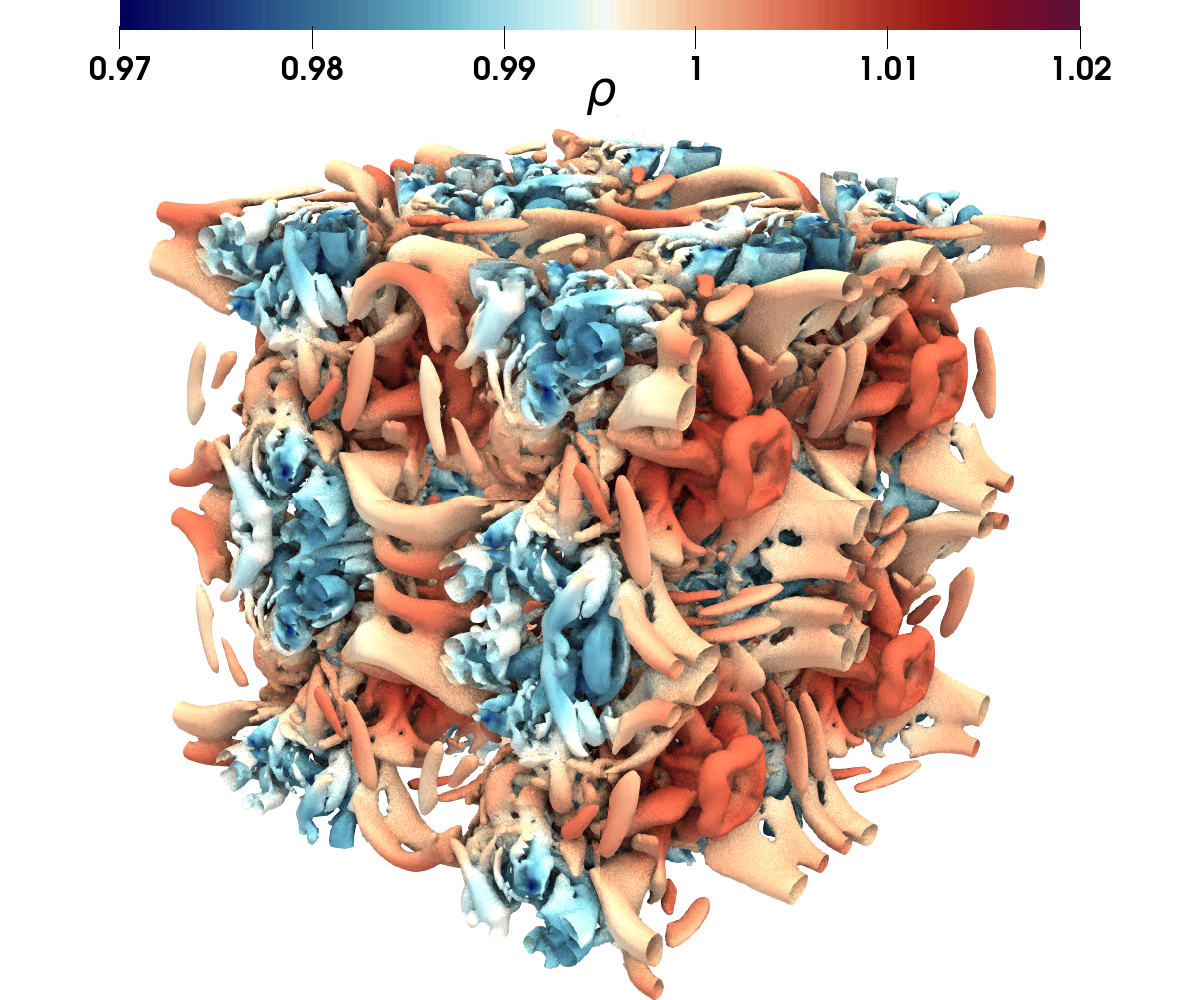}}}
    ~
    \subfloat[Boltzmann--BGK]{\adjustbox{width=0.48\linewidth, valign=b}{\includegraphics[width=\textwidth]{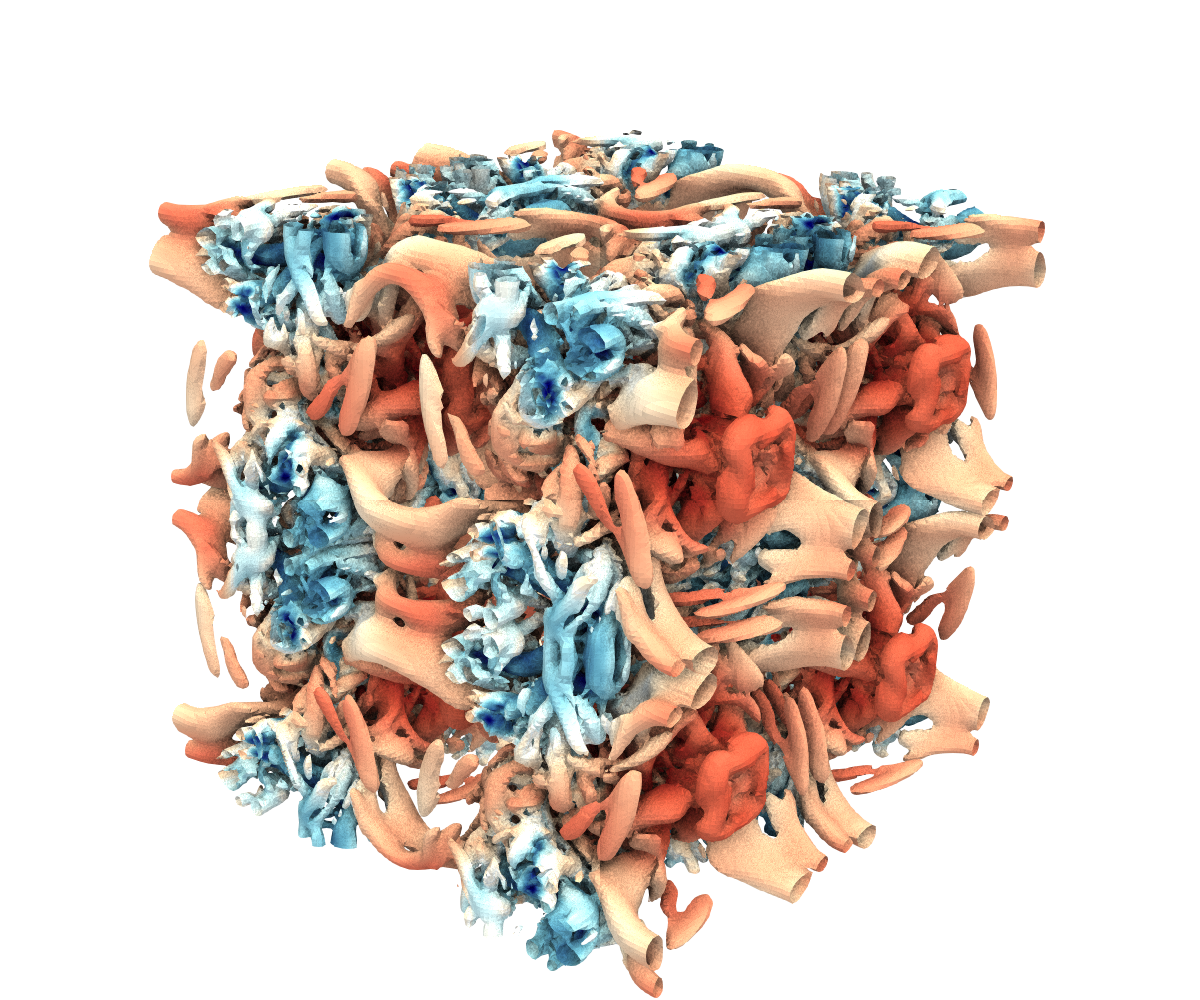}}}
    \caption{\label{fig:tgv_qcrit} Isosurfaces of Q-criterion colored by density at $t = 10$ for the compressible Taylor--Green vortex at $Re = 1600$ computed using a $\mathbb P_3$  approximation.}
\end{figure}

While the application of the Boltzmann to rarefied gas dynamics is quite typical, its use for more complex continuum flows has been severely limited. Due to the improved efficiency and accuracy of the numerical approach, the simulation of transition to turbulence for a three-dimensional compressible Taylor--Green Vortex at $Re = 1600$ was carried out. To the authors' knowledge, this was the first simulation of a three-dimensional turbulent flow performed by directly solving the Boltzmann equation. It was shown that the molecular gas dynamics equations could accurately predict nonlinear flow phenomena such as transition to turbulence consistently with the hydrodynamic equations, as shown through the enstrophy-based dissipation and turbulent kinetic energy spectra in \cref{fig:tgv_profiles}. Furthermore, nearly identical approximation of the turbulent flow structures was obtained with the Boltzmann--BGK equation as with the Navier--Stokes equations, shown through the Q-criterion isosurfaces in \cref{fig:tgv_qcrit}.

\begin{figure}[h!]
    \centering
    \subfloat[Reference] {
    \adjustbox{width=0.5\linewidth,valign=b}{\includegraphics[width=\textwidth]{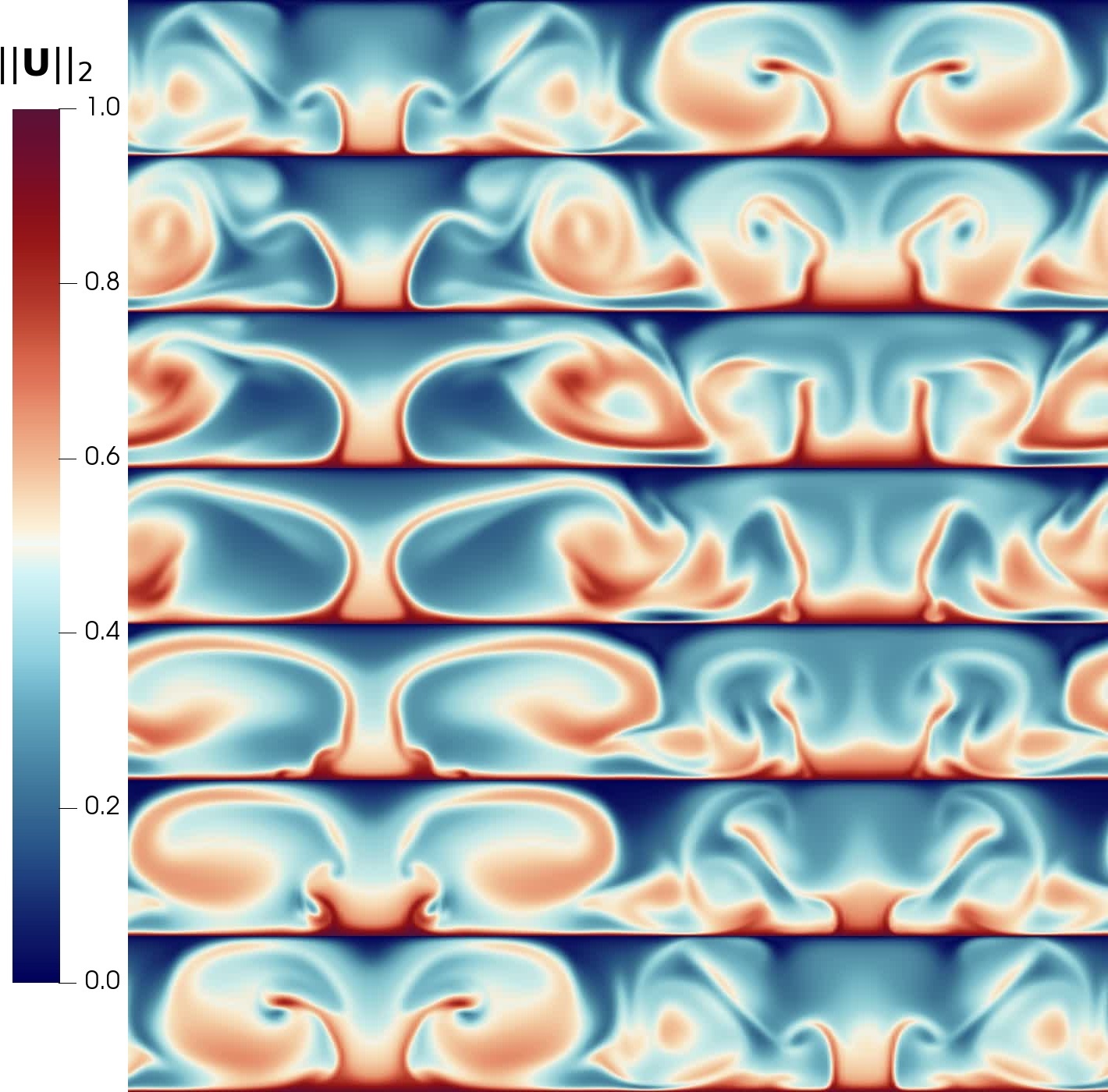}}}
    \subfloat[Boltzmann--BGK] {
    \adjustbox{width=0.44\linewidth,valign=b}{\includegraphics[width=\textwidth]{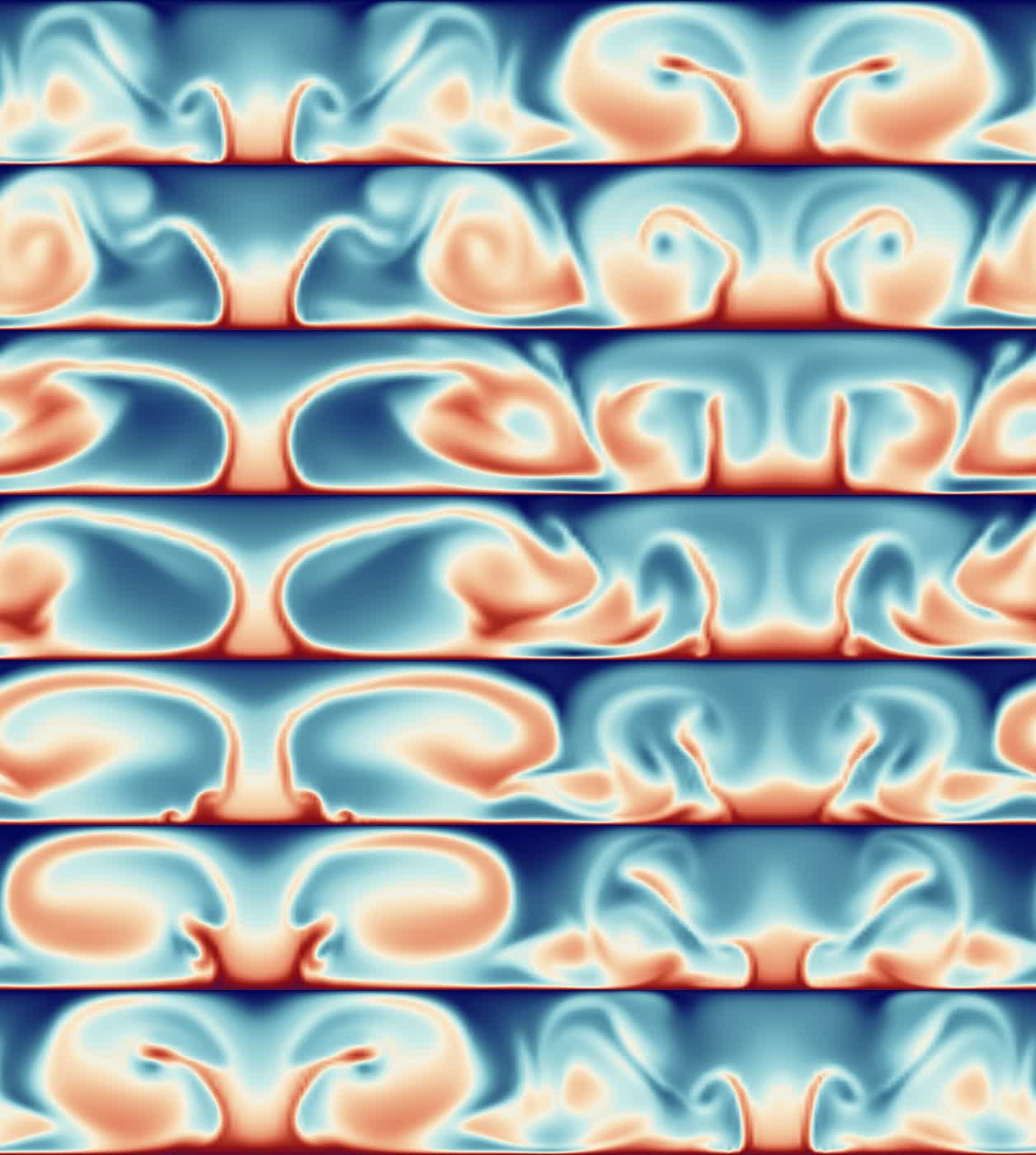}}}
    \caption{\label{fig:taylorcouette_velocity} Contours of velocity magnitude for the three-dimensional transitional Taylor--Couette flow problem at $t = 22$ on angularly equispaced cross-sections from $\theta = 0$ (top) to $\theta = \pi$ (bottom) computed using a $\mathbb P_5$ approximation.}
\end{figure}

\begin{figure}[h!]
    \centering
    \subfloat[Reference] {
    \adjustbox{width=0.5\linewidth,valign=b}{\includegraphics[width=\textwidth]{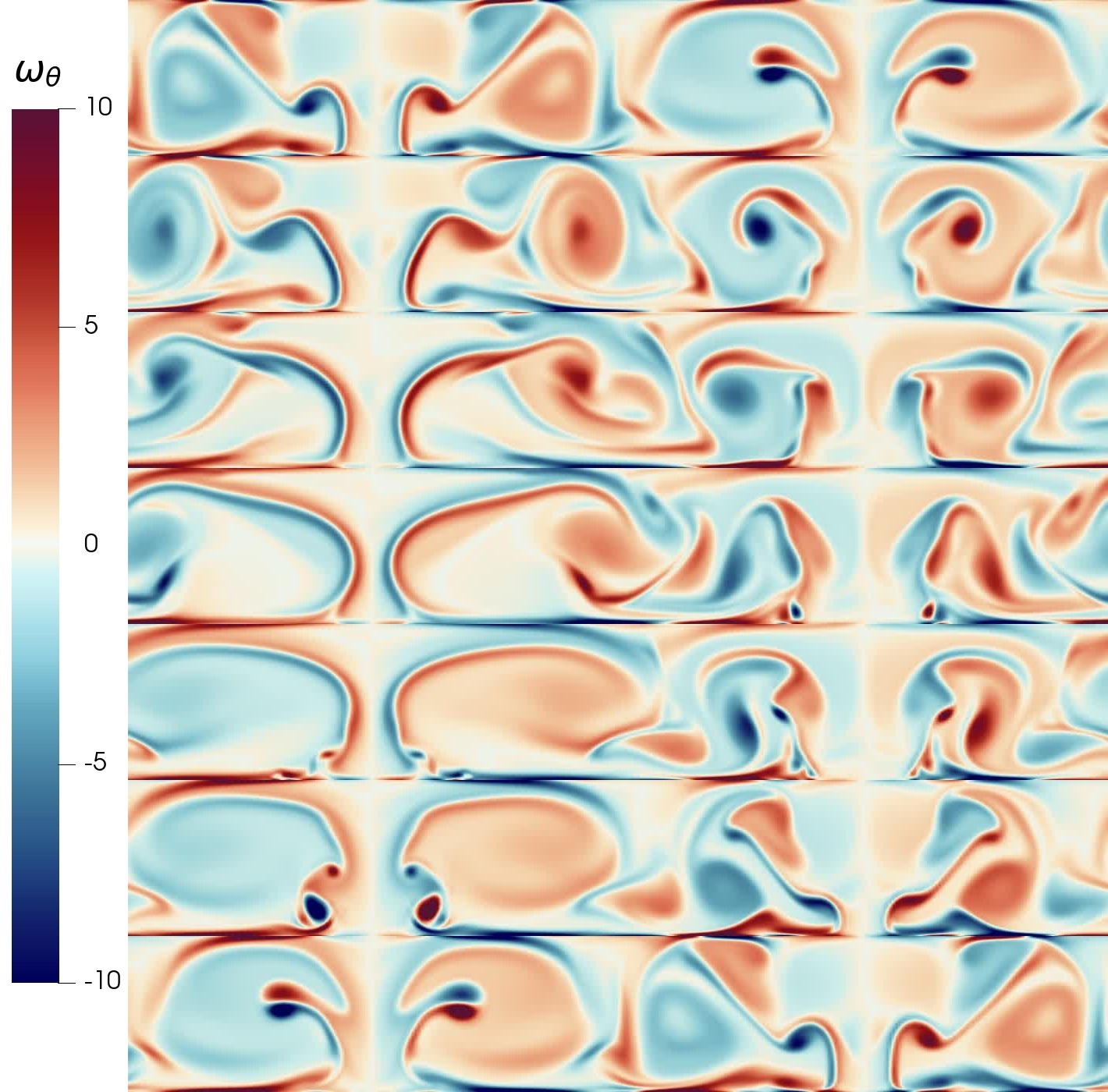}}}
    \subfloat[Boltzmann--BGK ] {
    \adjustbox{width=0.44\linewidth,valign=b}{\includegraphics[width=\textwidth]{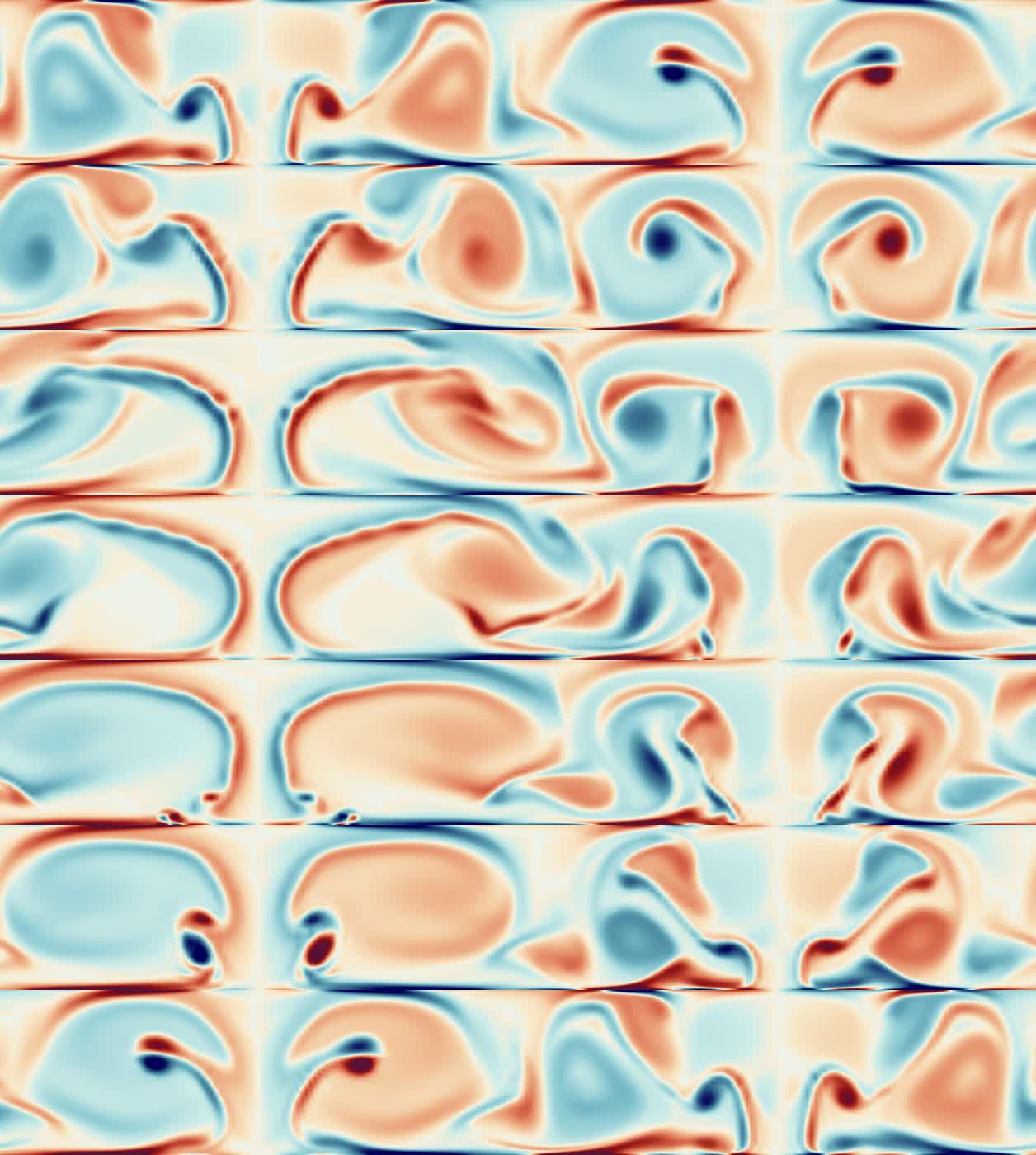}}}
    \caption{\label{fig:taylorcouette_vorticity} Contours of vorticity along the cross-section normal direction for the three-dimensional transitional Taylor--Couette flow problem at $t = 22$ on angularly equispaced cross-sections from $\theta = 0$ (top) to $\theta = \pi$ (bottom) computed using a $\mathbb P_5$ approximation.}
\end{figure}

Although the Taylor--Green vortex demonstrates that the Boltzmann--BGK approach can accurately predict nonlinear flow phenomena such as transition to turbulence, the lack of wall-fluid interactions in the problem greatly reduces the complexity of the flow. For the Boltzmann equation, the correct choice and validity of wall boundary conditions is still somewhat of an open problem. A comprehensive validation of the effects of wall boundary conditions on momentum transfer for complex wall-bounded fluid flows was performed by the authors \citep{Dzanic2023b}, with one of the numerical experiments evaluating these effects on the three-dimensional transitional Taylor--Couette flow \citep{Taylor1923}. For this problem, the flow is driven by concentric rotating cylinders, with the initially laminar flow transitioning to a turbulent state \citep{Wang2021}. Much like with the Taylor--Green vortex, this was, to the authors' knowledge, the first simulation of a three-dimensional wall-bounded turbulent flow performed by directly solving the Boltzmann equation, which was obtained using $4.7$ billion degrees of freedom with approximately a 36 times larger computational cost than the Navier--Stokes approach. It was seen that the shear-induced transition to turbulence could be accurately resolved by the Boltzmann--BGK approach, with a comparison of the velocity and vorticity contours shown in \cref{fig:taylorcouette_velocity} and \cref{fig:taylorcouette_vorticity}, respectively. Notably, it was found that to resolve these effects consistently with the Navier--Stokes equations, accurate prediction of particle transport was significantly more important than than accurate prediction of particle collision, i.e., a highly-resolved spatial domain was much more important than a highly-resolved velocity domain. These findings suggest that high-order schemes may be especially well-suited for simulating complex flows via the Boltzmann equation and that directly solving the Boltzmann equation can be performed at a reasonable cost compared to the Navier--Stokes equations as a result of the relatively few degrees of freedom necessary in the velocity domain. 


\begin{figure}[h!]
    \centering
    \subfloat[Pressure coefficient]{\adjustbox{width=0.48\linewidth, valign=b}{    \begin{tikzpicture}[spy using outlines={rectangle, height=3cm,width=2.3cm, magnification=3, connect spies}]
		\begin{axis}[name=plot1,
		    axis line style={latex-latex},
		    axis x line=left,
            axis y line=left,
            clip mode=individual,
		    xlabel={$x/c$},
		    xtick={0,0.2,0.4,0.6,0.8,1},
    		xmin=0,
    		xmax=1,
    	    x tick label style={
        		/pgf/number format/.cd,
            	fixed,
            	fixed zerofill,
            	precision=1,
        	    /tikz/.cd},
    		ylabel={$C_p$},
    		ytick={-3, -2, -1, 0, 1, 2},
    		ymin=-3,
    		ymax=2,
    		y dir=reverse,
    		y tick label style={
        		/pgf/number format/.cd,
            	precision=1,
        	    /tikz/.cd},
    		legend style={at={(0.97, 0.97)},anchor=north east,font=\small,nodes={scale=1, transform shape}},
    		legend cell align={left},
    		style={font=\normalsize},	
                scale = 0.85]


      

			\addplot[color=black, style={ultra thick, dotted}]
				table[x=x,y=cp,col sep=comma,unbounded coords=jump]{./figs/data/sd7003_ns_p4_cp.csv};
    		\addlegendentry{Navier--Stokes}
      
			\addplot[color=red!80!black, style={very thick}]
				table[x=x,y=cp,col sep=comma,unbounded coords=jump]{./figs/data/sd7003_bgk_p4_cp.csv};
    		\addlegendentry{Boltzmann--BGK}

		\end{axis} 		
	\end{tikzpicture}}}
    ~
    \subfloat[Skin friction coefficient]{\adjustbox{width=0.48\linewidth, valign=b}{    \begin{tikzpicture}[spy using outlines={rectangle, height=3cm,width=2.3cm, magnification=3, connect spies}]
		\begin{axis}[name=plot1,
		    axis line style={latex-latex},
		    axis x line=left,
            axis y line=left,
            clip mode=individual,
		    xlabel={$x/c$},
		    xtick={0,0.2,0.4,0.6,0.8,1},
    		xmin=0,
    		xmax=1,
    	    x tick label style={
        		/pgf/number format/.cd,
            	fixed,
            	fixed zerofill,
            	precision=1,
        	    /tikz/.cd},
    		ylabel={$C_f$},
    		ytick={-0.02, -0.01, 0, 0.01, 0.02},
    		ymin=-.02,
    		ymax=.02,
    		y tick label style={
        		/pgf/number format/.cd,
            	precision=1,
        	    /tikz/.cd},
    		legend style={at={(0.97, 0.97)},anchor=north east,font=\small},
    		legend cell align={left},
    		style={font=\normalsize},	
                scale = 0.85]

			\addplot[color=black, style={ultra thick, dotted}]
				table[x=x,y=cf,col sep=comma,unbounded coords=jump]{./figs/data/sd7003_ns_p4_cf.csv}; 
   	
			\addplot[color=red!80!black, style={very thick}]
				table[x=x,y=cf,col sep=comma,unbounded coords=jump]{./figs/data/sd7003_bgk_p4_cf.csv} ;

		\end{axis} 		
	\end{tikzpicture}}}
    \caption{\label{fig:sd7003_profiles} Surface pressure coefficient (left) and suction-side skin friction coefficient (right) for the $Re = 60,000$ SD7003 airfoil case computed with the Navier--Stokes equations (black) and the Boltzmann--BGK equation (red) using a $\mathbb P_4$ approximation. }
\end{figure}
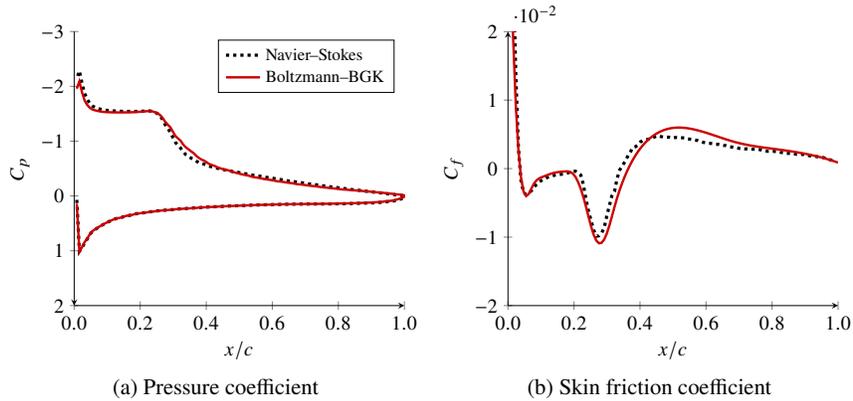

\begin{figure}[h!]
    \centering
    \subfloat[Q-criterion]{\adjustbox{width=0.48\linewidth, valign=b}{\includegraphics[width=\textwidth]{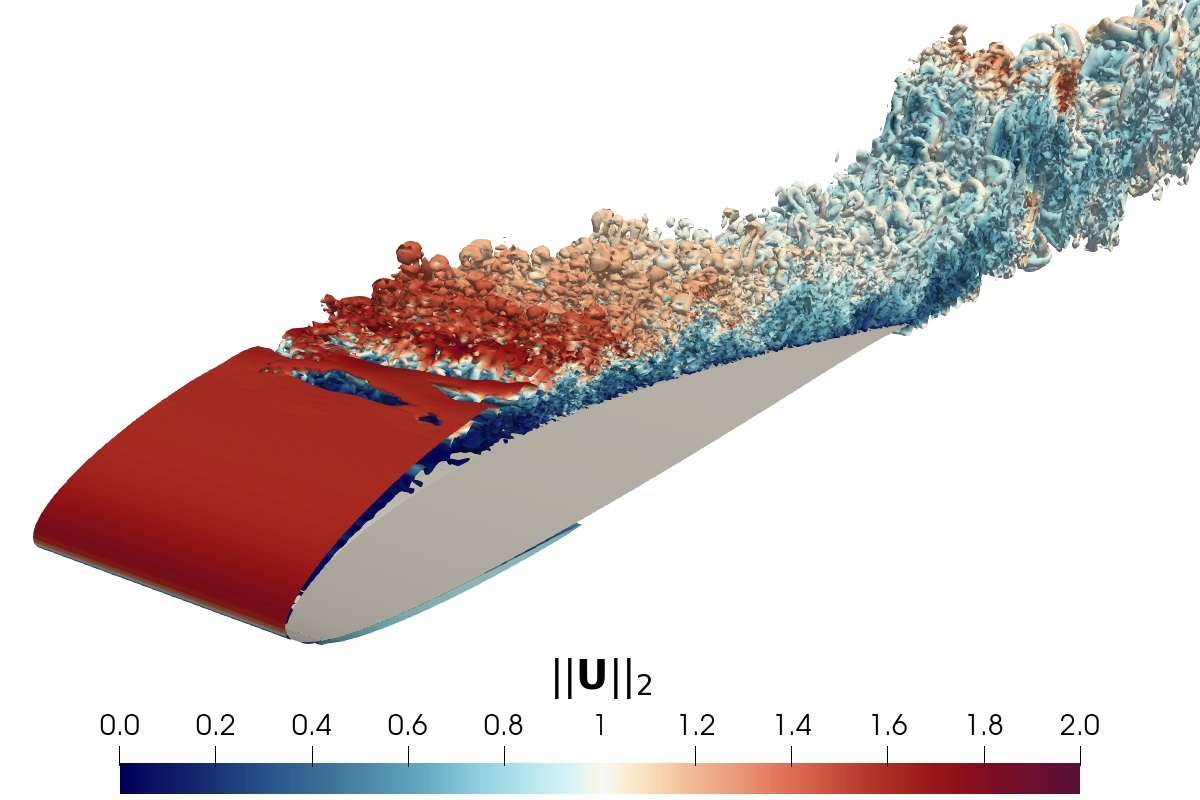}}}
    ~
    \subfloat[Streamwise velocity]{\adjustbox{width=0.48\linewidth, valign=b}{\includegraphics[width=\textwidth]{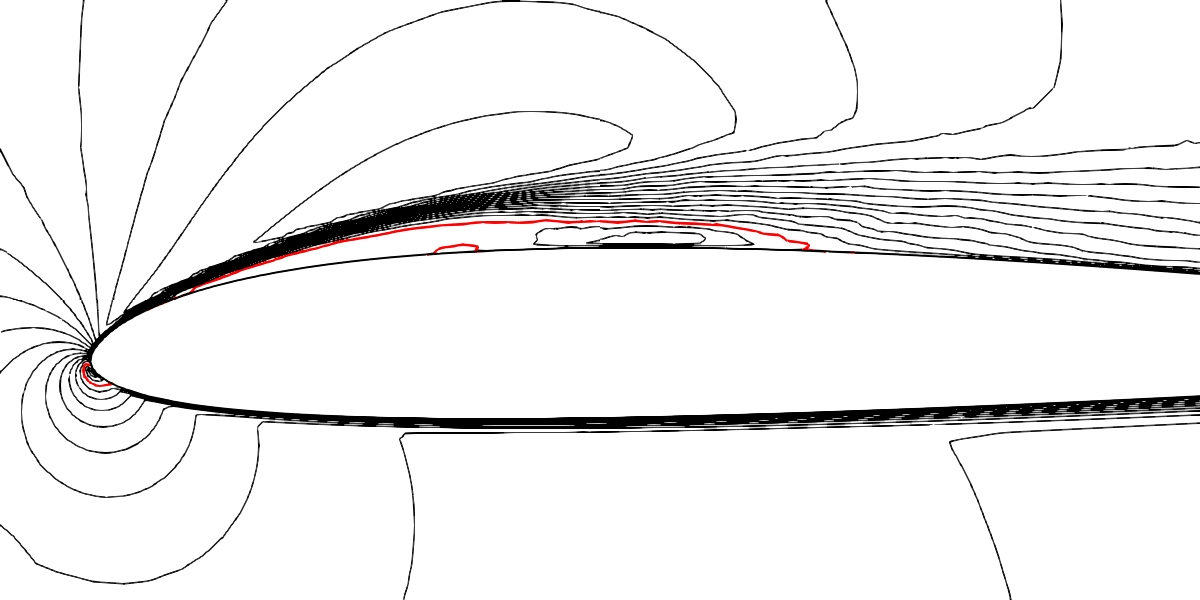}}}
    \caption{\label{fig:sd7003_iso} Instantaneous Q-criterion isosurfaces colored by velocity magnitude (left) and average streamwise velocity contours, equispaced on the range [-0.2, 1.5], (right) for the $Re = 60,000$ SD7003 airfoil case computed with the Boltzmann--BGK approach using a $\mathbb P_4$ approximation.}
\end{figure}

A more rigorous validation of the effects of wall-boundary conditions was performed through the simulation of the flow around an SD7003 airfoil at a Reynolds number of $Re = 60,000$ and angle of attack of $\alpha = 8^{\circ}$, which was obtained using $88.3$ billion degrees of freedom with approximately a 62 times larger computational cost than the Navier--Stokes approach. This case exhibits a laminar separation bubble that subsequently transitions to turbulence and reattaches into a turbulent boundary layer and wake \citep{Garmann2012}, which makes it highly sensitive to the flow conditions and difficult to accurately resolve. A comparison of the average surface pressure coefficient and 
skin friction coefficient as predicted by the Boltzmann--BGK approach and the Navier--Stokes approach is shown in \cref{fig:sd7003_profiles}. Good agreement between the two approaches was seen, both for the surface pressure coefficient and skin friction coefficient. Furthermore, a visualization of the instantaneous Q-criterion and average streamwise velocity contours as predicted by the Boltzmann--BGK approach is shown in \cref{fig:sd7003_iso}. The laminar separation bubble and the transition of the shear layer can be clearly seen through the Q-criterion isosurfaces and velocity contours, which showcases complex flow phenomena which can be accurately resolved by the Boltzmann--BGK approach. 

\begin{figure}[h!]
    \centering
    \subfloat[Surface slip velocity]{\adjustbox{width=0.48\linewidth, valign=b}{\includegraphics[width=\textwidth]{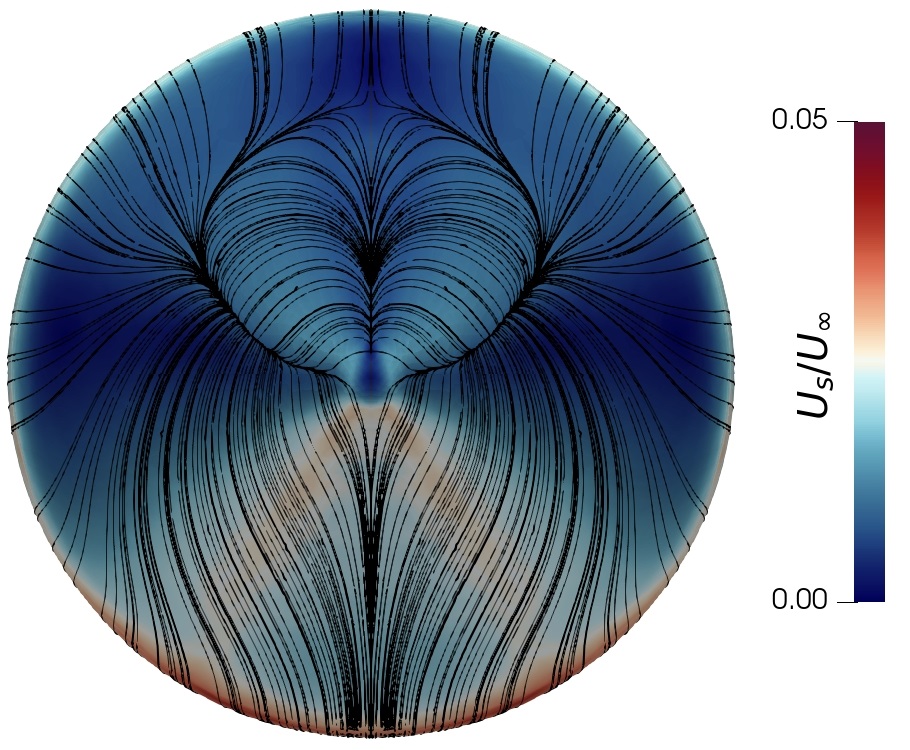}}}
    ~
    \subfloat[Density cross-section]{\adjustbox{width=0.48\linewidth, valign=b}{\includegraphics[width=\textwidth]{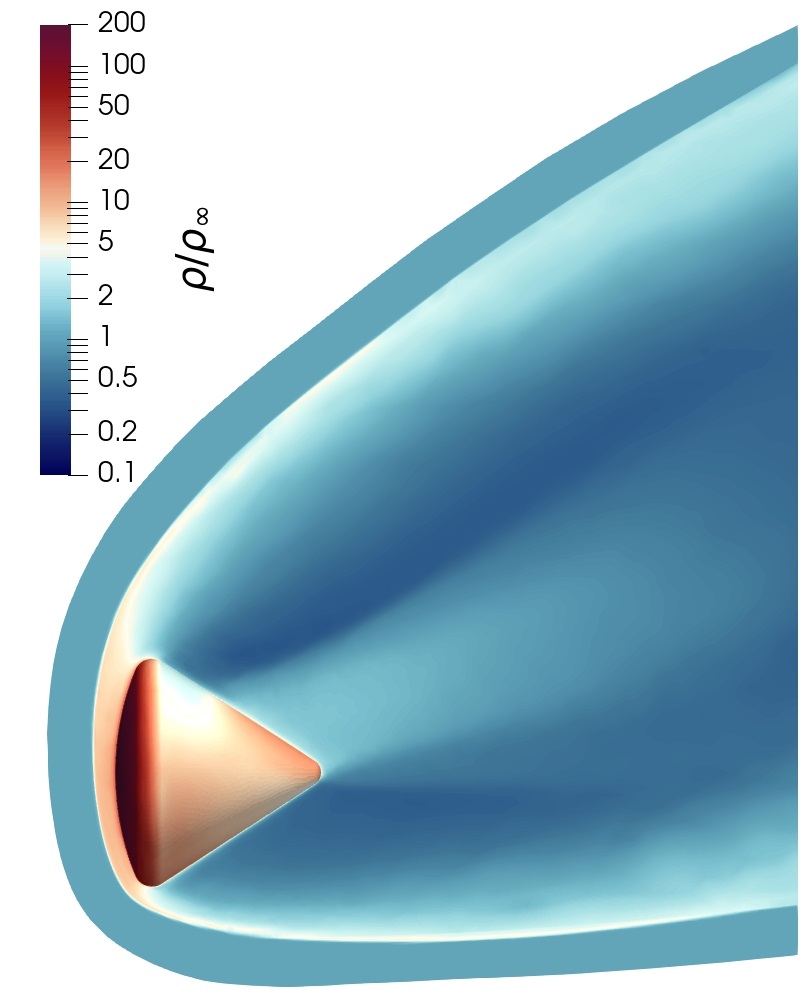}}}
    \caption{\label{fig:apollo} Surface slip velocity contours overlaid with slip velocity streamlines (left) and density contours on the cross-section $z = 0$ (right) for the Mach 22.7 Apollo re-entry capsule case at $Re = 43,000$ computed with the Boltzmann--BGK approach using a $\mathbb P_3$ approximation. }
\end{figure}

While the previous experiments focused on flows in the low Mach regimes, the use of the Boltzmann equation for high-speed flows also offers some notable advantages, such as the ability to directly resolve shock structures and strong aerothermodynamic effects. To demonstrate the ability of the Boltzmann--BGK approach for high-speed flows, we consider the flow around spacecraft in atmospheric re-entry conditions. In particular, the flow around a three-dimensional AS-202 Apollo capsule at a Mach number of 22.7 and a Reynolds number of 43,000 was simulated, which is, to the authors' knowledge, the first high-order simulation of a three-dimensional hypersonic flow obtained by the direct solution of the Boltzmann equation. The surface slip velocity contours/streamlines and a cross-section of the contours of density are shown in \cref{fig:apollo}. The strong detached bow shock can be clearly seen in the density contours which is resolved by the high-order numerical scheme \textit{without any ad hoc numerical shock capturing approach}. Furthermore, the wall slip velocity shows the local non-equilibrium nature of the flow, with complex unsteady behavior at the wall that would not be accurately predicted with the governing equations of continuum fluid dynamics.

\section{Conclusion}\label{sec:conclusion}
We present a brief overview of a numerical approach for simulating complex flows through directly solving the Boltzmann equation augmented with the BGK collision model. Through the combination of highly-efficient high-order spatial discretizations, discretely-conservative velocity models, and massively-parallel GPU computing, the approach allows for the simulation of complex three-dimensional flows ranging from rarefied microchannels to high-speed re-entry vehicles which were previously intractable. It was shown that complex nonlinear flow phenomena such as transition to turbulence as well as thermodynamic non-equilibrium effects could be accurately computed through this molecular gas dynamics approach. Furthermore, these simulations could be performed at a reasonable computational cost in comparison to the Navier--Stokes equations. The results of this work presents opportunities for entirely novel perspectives and approaches for complex flow problems, such as the analysis and study of fundamental flow phenomena through the evolution of a phase space distribution function and the use of a unified numerical framework for simulating flow problems that span a variety of flow regimes.

\bibliographystyle{unsrtnat}
\bibliography{references}
\end{document}